\documentclass[runningheads]{llncs}
\usepackage{times}
\usepackage{soul}
\usepackage{wrapfig}
\usepackage{url}
\usepackage[hidelinks]{hyperref}
\usepackage[utf8]{inputenc}
\usepackage[small]{caption}
\usepackage{graphicx}
\usepackage{caption}
\usepackage{booktabs}
\usepackage{algorithm}
\usepackage{algorithmic}
\usepackage[switch]{lineno}
\usepackage{comment}
\usepackage{hyperref}
\usepackage{color}
\usepackage{graphicx}
\usepackage{comment}
\begin{document}
\title{Lumos in the Night Sky: AI-enabled Visual Tool for Exploring Night-Time Light Patterns}
\titlerunning{Lumos in the Night Sky}
%
\author{Jakob Hederich \inst{1} \and
Shreya Ghosh\inst{2} \and
Zeyu He\inst{2} \and Prasenjit Mitra\inst{1,2}}
\authorrunning{Hederich and Ghosh et al.}
%
\institute{L3S Research Center,  Leibniz University, Hannover, Germany \and
College of IST, Pennsylvania State University, USA
\email{jakob.hederich@stud.uni-hannover.de},
\email{\{shreya,zeyuhe,pmitra\}@psu.edu}}
\maketitle              
\begin{figure}
  \centering
  \includegraphics[width=\textwidth]{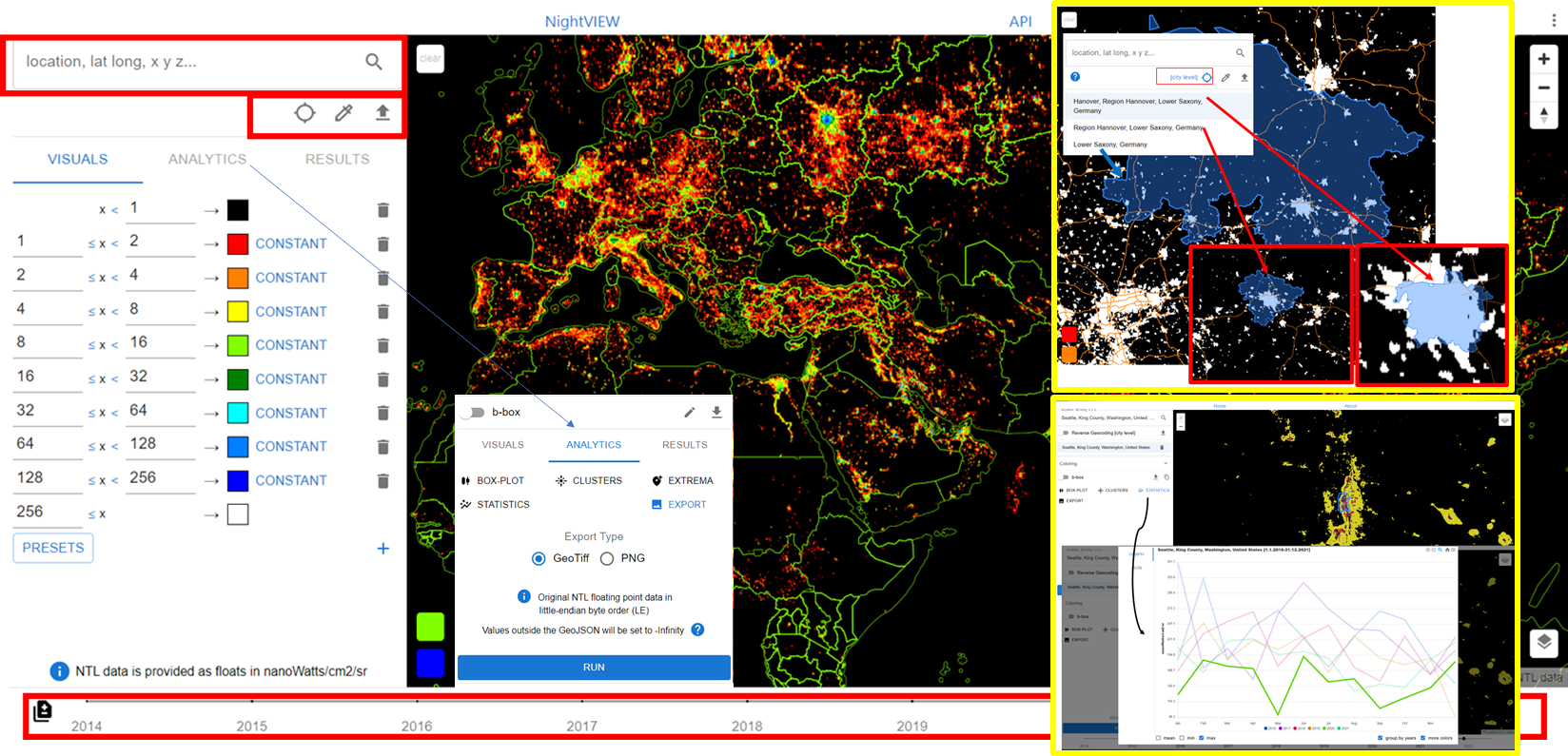}
  \captionof{figure}{{\small NightPulse: Red boxes depict spatial and temporal selection, and features (reverse geocoding, finding extrema, compare), 
  analytics box offers different computing features including clustering, temporal trend and export based on any selection of region and time-span. 
  On the top-right, we show the reverse-geocoding feature, where the region (Hannover, Lower Saxonomy) is selected based on zoom level (city, state, etc.). End-users can select/ define any range of NTL values for visualization. In the bottom-right, the user compares
  the NTL usage in 2020 and finds it to be lower than that in 2016-2019 in Seattle because people stayed home during lockdown.}}
  \label{fig:teaser}
\end{figure}

\begin{abstract}
We introduce NightPulse, an interactive tool for Night-time light (NTL) data visualization and analytics, which enables researchers and stakeholders to explore and analyze NTL data with a user-friendly platform. Powered by efficient system architecture, NightPulse supports image segmentation, clustering, and change pattern detection to identify urban development and sprawl patterns. It captures temporal trends of NTL and semantics of cities, answering questions about demographic factors, city boundaries, and unusual differences.\href{https://drive.google.com/file/d/1q5qlSF4vR7fPnmfhlQQw8WooMqnbLGgZ/view?usp=sharing}{ \textcolor{red}{Demo link}}

\keywords{Visualization  \and Night-time light (NTL) \and pattern mining.}
\end{abstract}
%
%
%
\section{Introduction}
Night-time light (NTL) data~\cite{viirsweb}, derived from satellite imagery~\cite{li2020harmonized}, has emerged as a critical resource for understanding human activity~\cite{li2020study,sanchez2022environmental,kyba2023citizen}, urban development~\cite{fang2022drives,chen2022potential,sanchez2022environmental}, and animal behavior~\cite{amichai2019artificial}. NTL data offers numerous advantages over traditional sources, such as consistent and continuous information and applicability in areas with limited data availability~\cite{li2016remote}. However, the effective analysis and interpretation of NTL data, necessitates overcoming several technical challenges including: (a) Data volume and accessibility: Downloading and managing  ~140GB in GeoTiff images of NTL monthly composite data (2015) from the NASA website. (b)Upscaling automated data analysis: Examining spatial patterns, such as urban growth and land use changes, and temporal patterns, such as annual or seasonal trends can be time-consuming. 
(c) Contextual data fusion: Integrating NTL data with other data, such as road networks, political maps, and population density, is challenging due to discrepancies in spatial and temporal resolutions, data formats, and calibration methods. (d) Customized visualization and user queries: 
Supporting high throughput end-user querying and intuitive visualizations for pattern detection is difficult without a comprehensive visualization tool. 

For instance, a few application questions are: \textbf{Q1:} How do urban sprawl and land use patterns visually manifest in NTL data over time? Can a visualization tool enable the comparison of NTL data from different regions or cities to assess their relative development, economic activity, or energy efficiency? \textbf{Q2:} How can a visualization tool help analyze and display the impact of major events, such as natural disasters, political changes/war, or economic shifts, on NTL data? \textbf{Q3:} How do cultural practices and festivities impact NTL data, particularly in regions with distinct religious or cultural events? For instance, a study~\cite{cultural} using NASA-NOAA satellite data showed Middle Eastern cities experienced a 50-100\% increase in brightness during Ramadan. This variation in lighting patterns was not solely determined by economic or commercial factors, but also reflected social and cultural identities. 

\par 
We introduce \textbf{NightPulse, \it{a semi-automatic, interactive visual analytics toolkit}},
which enables real-time NTL data exploration and extraction of insights. NightPulse surpasses existing tools (Kepler.gl\footnote{\url{https://kepler.gl/}}, NASA
Worldview\footnote{\href{https://worldview.earthdata.nasa.gov/?v=-88.14009998012352,-54.40325134172471,201.17770368044216,81.44049865827529&l=Reference_Labels_15m(hidden),Reference_Features_15m(hidden),Coastlines_15m(hidden),VIIRS_SNPP_DayNightBand_ENCC(hidden),VIIRS_Black_Marble,VIIRS_SNPP_CorrectedReflectance_TrueColor(hidden),MODIS_Aqua_CorrectedReflectance_TrueColor(hidden),MODIS_Terra_CorrectedReflectance_TrueColor&lg=false&tr=explore_the_earth_at_night&t=2016-01-01-T00\%3A00\%3A00Z}{NASA Worldview}}, tool by International Dark Sky association\footnote{\url{https://www.darksky.org/our-work/conservation/idsp/finder/}}, GeoTime~\cite{kapler2004geo}, TrajAnalytics~\cite{zhao2016trajanalytics}) in functionality and versatility. 
\section{NightPulse: Key visualization and computing features}
NightPulse is a flexible and user-friendly platform 
with customizable time and region selection, overlays, OSM map layer integration, pipette and extrema features, reverse geocoding, and image segmentation with thresholding ( Fig.~\ref{fig:teaser}). 
Users can integrate its output into workflows, precisely analyze NTL data, and identify spatial patterns/trends.


   \begin{figure}[htb]
       \centering
       \includegraphics[width=1.0\textwidth]{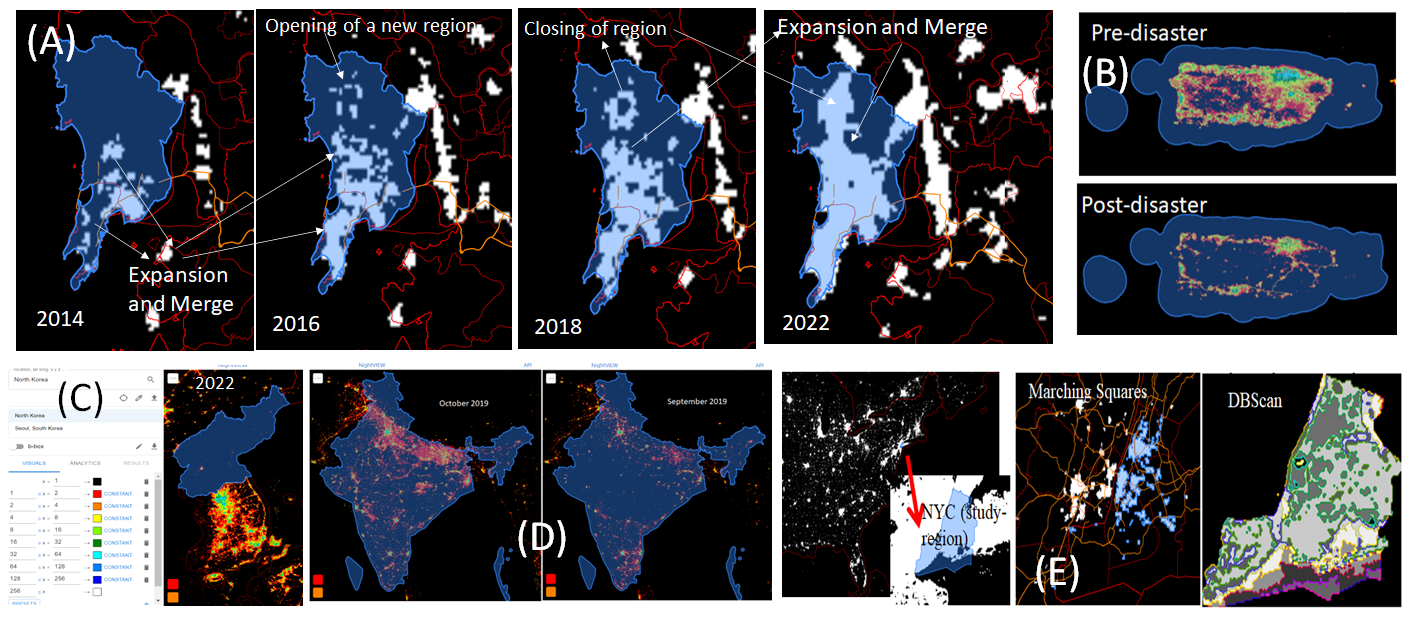}
       \caption{(A) Mumbai's Urban Sprawl Evolution (2014-2022), (B) Hurricane Maria's Impact on Puerto Rico (2017), (C) North and South Korea's NTL Distribution, (D) Diwali lights up India, and (E) Marching Squares and NightPulse-DBScan Algorithm applied to NY}
       \label{fig:interpret}
   \end{figure}
\noindent \textbf{Clustering:} NightPulse deploys the Marching Squares~\cite{maple2003geometric} algorithm for generating contours from the NTL map. 
The algorithm calculates the NTL value for each grid cell independently. Subsequently, a cell index is computed by comparing the contour levels with the NTL intensity values at the cell corners. Finally, a pre-constructed lookup table is employed to describe the output geometry for each cell. Algorithm~2 incorporates the temporal information with  DBSCAN~\cite{schubert2017dbscan}. 
By incorporating temporal information and intensity-based filtering, the algorithm enables the discovery of clusters that exhibit both spatial proximity and temporal consistency. 
Fig.~\ref{fig:interpret}(E) reveals insights (six major clusters, represented by distinct color boundaries) by deploying the NightPulse-DBScan algorithm (right image) and Marching squares algorithm (left) on NYC images (2014-2021). The results highlight consistently high night-time light intensity in Manhattan, seasonal variations in areas like 
the Rockefeller Center, and long-term trends of increasing light intensity in peripheral neighbourhoods (Staten Island or the far reaches of Brooklyn and Queens), indicating urban expansion and infrastructure development.
\\
\textbf{Urban sprawl pattern detection:} 
 NightPulse uses mathematical morphology~\cite{haralick1987image} to analyze and process images based on their shapes to detect urban sprawl. 
 Erosion, dilation, opening, and closing operations are applied to quantify shrink, merge, split, and expand patterns (Algo~1). Urban sprawl patterns, such as shrink, merge, expand, and split, can be measured and quantified by analyzing the regions in the transformed images. Structural changes in urban areas occur due to population dynamics, booming economies, or improved connectivity. Fig.~\ref{fig:interpret}(A) captures Mumbai's growth pattern through NTL data from 2014 to 2022. NightPulse helps identify and quantify these different types of changes, which cannot be achieved by simply comparing two images.
\begin{figure}[htb]
    \centering
    \includegraphics[width=1.0\textwidth]{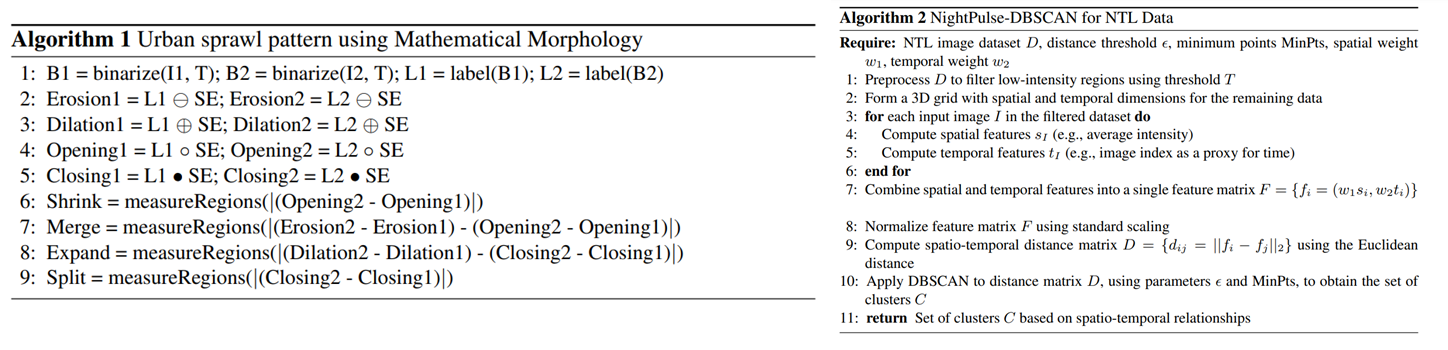}
   \end{figure}

   \begin{figure}[htb]
       \centering       \includegraphics[width=1.0\textwidth,height=3.5cm]{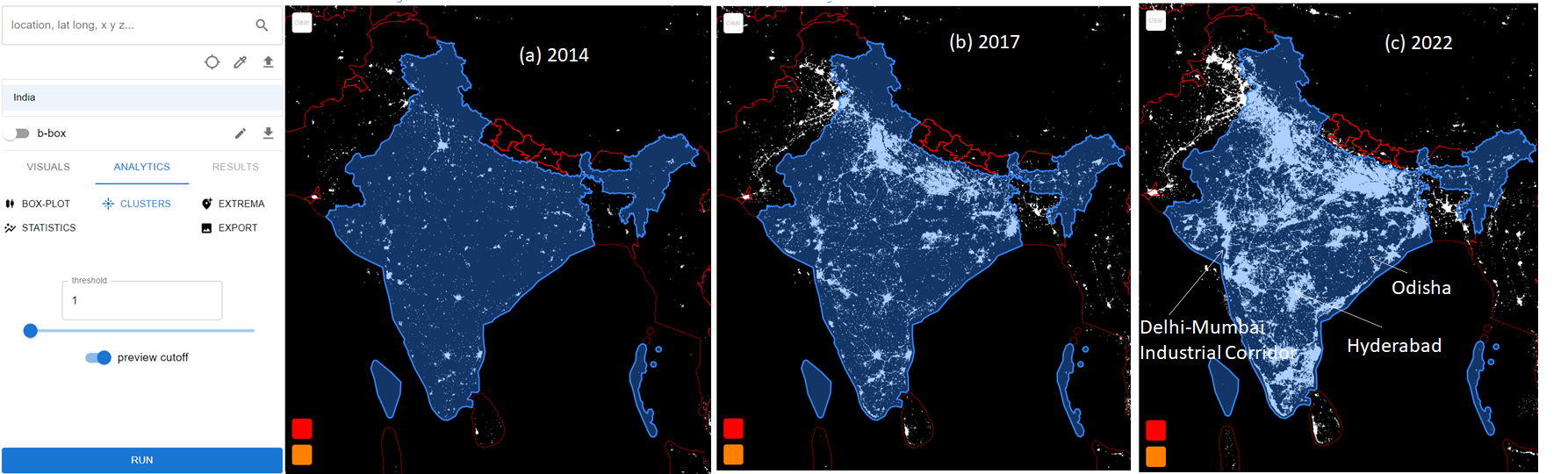}
       \caption{Urbanization, Infrastructure Development, and Economic Growth. India:'14,'17, and '22}
       \label{fig:indiachange}
   \end{figure}

\textbf{Data interpretation capacity:} {\it (1) Propaganda and accuracy of government statistics:} Pyongyang and Seoul 
have vastly different night-time lighting patterns ( Fig.~\ref{fig:interpret}(C)). North Korea's centrally planned economy, energy shortages, and political isolation have 
created sparse and dim night-time lighting; South Korea's bustling economy, modern infrastructure, and global integration have created a bright and extensive lighting pattern in Seoul.
{\it (2) Economic Development:}
India's NTL data (See Fig.~\ref{fig:indiachange}) from 2014, 2017, and 2022, shows urban expansion, infrastructure development, and economic growth, e.g., expansion in the Delhi-NCR region, Bengaluru, and Hyderabad
metropolitan areas, infrastructure development (such as highways, airports, and industrial zones, development of industrial corridors like the Delhi-Mumbai Industrial Corridor); economic growth (correlated to India's overall economic development and rural electrification); spatial inequality (e.g.,
Odisha's lagging economic growth and infrastructure development).(3) {\it Assessing disaster impact:}
E.g., power outages and the pace of recovery. Fig.~\ref{fig:interpret}(B) shows the impact of Hurricane Maria on Puerto Rico after the hurricane made landfall on September 16, 2017. This is valuable for disaster response, resource allocation, and assessing the effectiveness of recovery efforts. 
We can see the socio-economic impacts of the Russia-Ukraine war on affected regions, e.g., NTL drop due to disruptions in human activity, infrastructure damage, or population displacement, as well NTL increases due to Diwali across India (Fig~\ref{fig:interpret}(D)).\\ \\ 
\noindent\textbf{Demonstration plan and conclusion}
We will demonstrate the following (and more): \\
\textbf{DQ1: Can NightPulse connect socioeconomic indicators, such as income and population density, and energy consumption patterns?} A1: Its GeoJSON upload and overlay features allow users to incorporate additional data layers, like income and population density, on top of the NTL map. 
Users can  correlate NTL and overlaid features, e.g., socioeconomic factors and energy consumption across the world at different times.\\ \textbf{DQ2: Can NightPulse identify natural disasters or monitor recovery efforts?} A2: 
NightPulse's compare and clustering identify natural disasters like earthquakes, hurricanes, or floods, provide insights into the pace of recovery and infrastructure rebuilding. \\
\textbf{DQ3: Identifying City Borders and Green Spaces}
Thresholding and light contours pinpoint city limits and areas in a city that are greener, either in terms of vegetation or eco-friendly practices or less green such as airports. Parks, urban forests, or neighborhoods have low NTL usage and airports high.\\ \textbf{DQ4: Recognizing Specific Map Elements from NTL Data} Can NightPulse distinguish map components like roads, railway networks, and Points of Interest (POIs), such as commercial centers and residential areas? A4: Major highways and other well-traveled routes often appear as bright lines connecting cities and regions. POIs exhibit unique NTL signatures that differentiate between varying types of land use, e.g., shopping malls, office complexes, or  distinct residential neighborhoods.

NightPulse is the first visual analytics toolkit that provides real-time answers and analyses to queries on NTL data posed by
researchers, government stakeholders, NGOs, and private sector organizations involved in urban planning, policy-making, and infrastructure development.
End-users can make
informed decisions and design better policies based on insights into human activity
patterns and urban development trends.

\end{document}